\documentclass[letterpaper, 10 pt, conference]{ieeeconf}  

\IEEEoverridecommandlockouts                              

\usepackage{graphics} 
\usepackage{amsmath} 
\usepackage{amssymb}  
\usepackage{cite}
\usepackage{graphicx}
\usepackage{booktabs}
\usepackage{hyperref}
\usepackage{url}
\usepackage{float}

\floatstyle{ruled}
\newfloat{algorithm}{htb}{loa}
\floatname{algorithm}{Algorithm}

\newcounter{algline}
\newlength{\algindent}

\newenvironment{algorithmic}{%
  \setcounter{algline}{0}%
  \begin{list}{}{%
    \setlength{\leftmargin}{2.6em}%
    \setlength{\labelwidth}{2.0em}%
    \setlength{\labelsep}{0.6em}%
    \setlength{\itemsep}{1pt}%
    \setlength{\parsep}{0pt}%
    \setlength{\topsep}{4pt}%
  }%
}{%
  \end{list}%
}

\newcommand{\State}[1]{%
  \stepcounter{algline}%
  \item[\footnotesize\thealgline:]%
  #1%
}
\newcommand{\Statex}[1][]{\item[] #1}
\newcommand{\Require}[1]{\item[\textbf{Require:}] #1}
\newcommand{\AlgReturn}[1]{\State{\textbf{return}\; #1}}
\newcommand{\algocomment}[1]{\hfill$\triangleright$\;\textit{\small #1}}
\newcommand{\In}[1]{\hspace{\algindent}#1}
\newcommand{\Inn}[1]{\hspace{2\algindent}#1}
\graphicspath{{figs/}}

\title{\LARGE \bf
Mixed-Integer Nonlinear Differentiable Predictive Control for Underground Pumped Hydro Energy Storage Systems
}

\author{Honghui Zheng$^{1}$, J\'an Boldock\'y$^{2}$, Yury Dvorkin$^{1}$ and  J\'an Drgo\v na$^{1}$
\thanks{$^{1}$ HZ, YD, and JD are with the Johns Hopkins University, Baltimore MD, US. ({hzheng39, ydvorki1, jdrgona1}@jh.edu).
}
\thanks{$^{2}$ JB is with the Slovak University of Technology in Bratislava, Slovakia. (jan.boldocky@stuba.sk)}
\thanks{This work is supported by Ralph O’Connor Sustainable Energy Institute. This research was also supported by the U.S. DOE, Office of Science, ASCR program under the Scientific Discovery through Advanced Computing (SciDAC) Institute “LEADS: LEarning-Accelerated Domain Science”. J.B. acknowledges the support provided by the Scientific Grant Agency of the Slovak Republic under the grant 1/0401/26.}
\thanks{\copyright~2026 IEEE. Personal use of this material is permitted. Permission from IEEE must be obtained for all other uses, in any current or future media, including reprinting/republishing this material for advertising or promotional purposes, creating new collective works, for resale or redistribution to servers or lists, or reuse of any copyrighted component of this work in other works. Accepted for publication in the 2026 65th IEEE Conference on Decision and Control (CDC).}%
}

\begin{document}

\maketitle
\thispagestyle{empty}
\pagestyle{empty}

\begin{abstract}
This paper extends Mixed-Integer Differentiable Predictive Control (MI-DPC) to multi-modal discrete decisions and nonconvex polynomial dynamics arising in Underground Pumped Hydro Energy Storage Systems (UPHES). A neural policy mapping problem parameters to continuous setpoints and integer mode selections via a Gumbel-Softmax layer is trained in a self-supervised manner by differentiating the expectation of the finite horizon control objective through the nonlinear dynamics model. Three methodological contributions enable this extension: a parallel differentiable simulator that preserves gradient magnitude, a Transformer encoder that captures long-range temporal dependencies, and a Gumbel-Softmax temperature annealing schedule that regularizes the combinatorial search. We demonstrate the framework on day-ahead scheduling of a UPHES, a large-scale mixed-integer optimal control problem with nonlinear unit performance curves and volume-head coupling. MI-DPC achieves only 1.6\% suboptimality relative to a piecewise mixed-integer quadratic programming baseline, while providing five orders of magnitude speedup in online scheduling time.
\end{abstract}

\section{INTRODUCTION}

Parametric mixed-integer nonlinear programs (MINLPs) with nonlinear dynamics and discrete mode decisions arise broadly in optimal
control problems, yet due to their NP-hard~\cite{bendotti2019complexity} nature remain computationally prohibitive for real-time solution, even with state-of-the-art solvers. This paper addresses a representative instance: day-ahead
scheduling of an Underground Pumped Hydro Energy Storage (UPHES) system~\cite{rehman2015pumped,blakers2021review}, where at each of
24 hourly steps the unit selects among three mutually exclusive modes (pump, idle, turbine) and determines continuous power setpoints
subject to polynomial unit performance curves (UPCs), cubic volume--head coupling, and time-varying electricity prices. 

Existing approaches sacrifice model fidelity for tractability. Mixed-integer programming (MIP) formulations replace nonlinear mappings with piecewise-linear approximations~\cite{toubeau2019non} or their chance-constrained extensions~\cite{toubeau2019chance}, and global affine surrogates~\cite{favaro2024neural}, enabling branch-and-bound solvers at the cost of systematic error that grows in high-curvature regions of the dynamics. Alternative strategies like Bayesian optimization~\cite{gobert2022parallel} or multi-fidelity simulators~\cite{favaro2024neural} improve solution accuracy but require computational budgets incompatible with real-time operation. A decision-focused learning framework~\cite{zheng2026accelerating} trains a neural network to guide the recursive linearization of the MINLP into differentiable quadratic programs, yet still solves an optimization problem at inference and inherits integer-variable errors from initializations.

From a control perspective, the scheduling problem is a parametric MINLP: given an initial state and price forecast, compute a state-input trajectory subject to nonlinear dynamics and integer constraints. Classical explicit model predictive control (MPC) enumerates polyhedral regions of the parameter space~\cite{bemporad2002explicit}, but the number of regions grows exponentially with horizon and constraint count~\cite{borrelli2017predictive}. Neural approximations of MPC learn the parameter-to-control map via supervised learning~\cite{karg2020efficient,hertneck2018learning}, but require data labels of pre-solved optimal instances, which can be prohibitively expensive to generate for MIPs~\cite{Cauligi2022,Bertsimas2022}.

To alleviate these issues, Differentiable Predictive Control (DPC)~\cite{drgovna2022differentiable,drgovna2024learning} introduced an MPC-inspired self-supervised training of a parametric control policy via automatic differentiation over known dynamics, enabling scalable offline pre-training and fast online inference. 
However, extending DPC to mixed-integer problems requires gradients through discrete mode decisions, although argmax is nondifferentiable almost everywhere. Gumbel-Softmax~\cite{jang2016categorical,maddison2016concrete} with the straight-through estimator~\cite{bengio2013estimating} provides a workaround: the forward pass uses hard one-hot decisions, while the backward pass follows a differentiable softmax surrogate. Boldo\-ck\'y et~al.~\cite{boldocky2025learning} used this approach to introduce MI-DPC for a thermal energy system with linear dynamics and MLP policies. Concurrent works extend MI-DPC to nonlinear dynamics with binary decisions for data center cooling~\cite{boldocky5764791data} and to battery dispatch with nondifferentiable degradation models~\cite{safarzadeh2026end}. Other self-supervised learning-to-optimize approaches for MINLP, including with feasibility guarantees~\cite{tang2024learning}, 
and mixed-integer MPC~\cite{ADAMEK2026103636}, exploit similar differentiable approximations to integer policies, yet
consider different settings from the MI-DPC framework here. A key consideration in the proposed Gumbel-Softmax policy representation is the temperature $\tau$~\cite{jang2016categorical}: high values encourage exploration but bias the relaxation, whereas low values sharpen decisions at the cost of weaker, noisier gradients and a greater risk of premature mode commitment, similar to relaxation collapse in differentiable architecture search~\cite{ichikawa2024controlling}. To the best of our knowledge, no prior work has studied this trade-off systematically in the context of MI-DPC.

\noindent\textbf{Contributions.} We extend the MI-DPC framework~\cite{boldocky2025learning} to UPHES scheduling, where polynomial UPCs and volume--head coupling introduce nonlinear, non-convex dynamics. To handle these challenges we introduce three methodological contributions: (i)~a parallel differentiable simulator that preserves training-time gradient magnitude through nonlinear dynamics; (ii)~a Transformer policy in place of an MLP; and (iii)~a two-stage temperature schedule for more stable discrete mode learning.
To facilitate adoption and reproducibility, we make the code open-source\footnote{\url{https://github.com/SOLARIS-JHU/MI-DPC-UPHES}}.

\section{PROBLEM FORMULATION}
\label{sec:problem}

We consider the day-ahead scheduling of an Underground Pumped Hydroelectric Storage (UPHES) unit operating as a price taker on the electricity market over a horizon of $N\!=\!24$ hours with hourly time steps $\Delta t\!=\!1\,\text{h}$. 
At each hour $t\!\in\!\mathcal{T}\!=\!\{1,\dots,N\}$ the unit selects one of three mutually exclusive operational modes: idle~($I$), turbine~($T$), or pump~($P$). The scheduling problem constitutes a mixed-integer nonlinear program (MINLP):

%
\begin{subequations}
\label{eq:minlp}
\begin{gather}
\underset{p_t^T,\,p_t^P,\,m_t}{\text{maximize}}
\;\sum_{t=1}^{N}\!\Big(\!(p_t^T\!+\!p_t^P)\,\lambda_t^{\text{DA}}
  -C_{\text{op}}(p_t^T\!+\!p_t^P)^2\Big) \label{eq:objective}\\[2pt]
\text{s.t.}\quad m_t\in\{-1,0,1\} \label{eq:mode}\\
p_{\min}^{T}(h_t)\,[m_t]_+\!\le p_t^{T}\le p_{\max}^{T}(h_t)\,[m_t]_+ \label{eq:bounds_T}\\
p_{\min}^{P}(h_t)\,[-m_t]_+\!\le p_t^{P}\le p_{\max}^{P}(h_t)\,[-m_t]_+ \label{eq:bounds_P}\\
q_t = f_{T}^{\text{UPC}}(p_t^{T},h_t)\,[m_t]_+
      + f_{P}^{\text{UPC}}(p_t^{P},h_t)\,[-m_t]_+ \label{eq:upc}\\
v_{\text{low},t} = v_{\text{low}}^{\text{init}}
                + \Delta t\sum_{s=1}^{t} q_s  \label{eq:vol_dyn}\\
v_{\text{low},t} = f^{\text{vol}}(h_t)  \label{eq:vol_head}\\
0\le v_{\text{low},t}\le v_{\max}  \label{eq:vol_bounds}\\
h_{\min}\le h_t\le h_{\max}  \label{eq:head_bounds}\\
v_{\text{low},N}\le v_{\text{low}}^{\text{target}} \label{eq:target}
\end{gather}
\end{subequations}
where $\lambda_t^{\text{DA}}$ is the day-ahead price and $C_{\text{op}}$ is a quadratic operational cost coefficient. The integer $m_t$ encodes the mode ($1$: turbine, $0$: idle, $-1$: pump), with $[m_t]_+ := \max(0,m_t)$ and $[-m_t]_+ := \max(0,-m_t)$. Power is signed as in~\cite{zheng2026accelerating}: $p_t^T\!\ge\!0$ is generation and $p_t^P\!\le\!0$ is consumption, so the pump bounds satisfy $p_{\min}^{P}(h_t)\!\le\!p_{\max}^{P}(h_t)\!\le\!0$ and the revenue in~\eqref{eq:objective} is negative when pumping. Constraints~\eqref{eq:bounds_T}--\eqref{eq:bounds_P} enforce head-dependent power limits, while~\eqref{eq:upc} obtains the flow rate $q_t$ from mode-specific unit performance curves (UPCs) $f_m^{\text{UPC}}(p,h)$, which are bivariate polynomials fitted to experimental Francis pump-turbine data~\cite{mercier2017provision}. The maximum reservoir capacity $v_{\max}$ is imposed by~\eqref{eq:vol_bounds}.

The lower-reservoir volume $v_{\text{low},t}$ evolves via cumulative flow~\eqref{eq:vol_dyn}, accumulating net flow over all preceding time steps. The nonlinear volume--head coupling~\eqref{eq:vol_head} arises from reservoir geometry~\cite{toubeau2019non}. Constraint~\eqref{eq:target} preserves long-term water balance. The polynomial UPCs, nonconvex volume--head map, and integer modes render~\eqref{eq:minlp} computationally challenging. Mixed-integer quadratic programming (MIQP) reformulations using either global linearization (MIQP-GL) or piecewise-bilinear approximations (MIQP-PW) are commonly employed, trading fidelity for tractability~\cite{zheng2026accelerating}.

Beyond the technical constraints, the operator faces two market-driven costs incorporated in the training loss (Section~\ref{sec:loss}): a \emph{system imbalance penalty} under a double-pricing settlement~\cite{vandezande2010well,bottieau2019very} (shortages penalized at a premium, surpluses compensated at a discount), and a \emph{target volume penalty} that monetizes any water left in the lower reservoir above the target in~\eqref{eq:target}, i.e., an end-of-horizon state-of-charge deficit, at the median day-ahead price.

\section{METHODOLOGY}
\label{sec:methodology}

\begin{figure*}
    \centering
\includegraphics[width=0.8\linewidth]{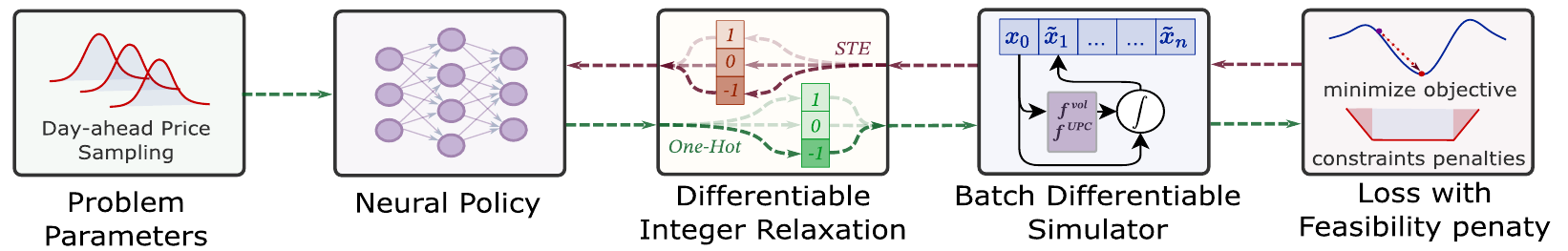}
\vspace{-0.1cm}
    \caption{MI-DPC pipeline with Softmax STE. Problem parameters are mapped by a neural policy to continuous power ratios and mode logits. A Gumbel-Softmax STE yields hard mode decisions, the differentiable simulator rolls out the nonlinear UPHES dynamics, and the Lagrangian loss combines negative ex-post profit with soft constraint penalties. Green arrows denote the forward pass; red arrows denote backward gradient flow through the STE and simulator.}
    \label{fig:midpc}
\end{figure*}

This section reformulates the MINLP~\eqref{eq:minlp} as a differentiable program within the DPC framework~\cite{drgovna2024learning,boldocky2025learning}. Nonlinear bounds~\eqref{eq:bounds_T}--\eqref{eq:bounds_P} and dynamics~\eqref{eq:upc}--\eqref{eq:vol_head} are handled in a differentiable simulator, using straight-through estimators for hard projections that block gradients. The target constraint~\eqref{eq:target} and market settlement terms enter the loss~\eqref{eq:total_loss}, while Section~\ref{sec:ste} introduces the integrality mechanism for $m_t$. Figure~\ref{fig:midpc} demonstrates the MI-DPC pipeline.

\subsection{Parametric Formulation}
\label{sec:parametric}

The MINLP~\eqref{eq:minlp} is parameterized by the initial hydraulic state and day-ahead price trajectory:
\begin{equation}
\boldsymbol{\xi}=[h^{\text{init}},\,v^{\text{init}},\,\lambda_1^{\text{DA}},\dots,\lambda_{N}^{\text{DA}}]^\top\!\in\mathbb{R}^{N+2}
\label{eq:xi}
\end{equation}
A neural policy $\pi_\theta : \mathbb{R}^{N+2} \to [0,1]^{N\times 2} \times \mathbb{R}^{N\times 3}$ maps $\boldsymbol{\xi}$ directly to continuous power ratios $\boldsymbol{u}_c\in[0,1]^{N\times2}$ and discrete mode logits $\boldsymbol{\ell}\in\mathbb{R}^{N\times3}$:
$
[\boldsymbol{u}_c,\,\boldsymbol{\ell}]^{\top} = \pi_\theta(\boldsymbol{\xi}).
$ 
Training minimizes the expected penalized loss over $\mathcal{P}_{\boldsymbol{\xi}}$:
\begin{equation}
\min_\theta\;\mathbb{E}_{\boldsymbol{\xi}\sim\mathcal{P}_{\boldsymbol{\xi}}}
\!\Big[\mathcal{L}\!\big(\pi_\theta(\boldsymbol{\xi});\,\boldsymbol{\xi}\big)\Big],
\label{eq:dpc_obj}
\end{equation}
where $\mathcal{L}$ combines negative profit and soft constraint penalties, discussed further in Section~\ref{sec:loss}.

\subsection{Neural Policy Architecture}
\label{sec:architecture}

The policy uses a Transformer encoder with two output heads. At each hour, the input token $\boldsymbol{z}_t:=[\bar{\lambda}_t,\,\bar{h},\,\bar{v}]^\top$ contains the min-max normalized price $\bar{\lambda}_t$ and repeated initial states $\bar{h},\,\bar{v}$. Tokens are projected by a small MLP, augmented with sinusoidal positional encodings, and processed over the full horizon by a multi-layer Transformer encoder~\cite{vaswani2017attention}. Two MLP heads then output the continuous power ratios $\boldsymbol{u}_c\!\in\![0,1]^{N\times2}$ via sigmoid and the mode logits $\boldsymbol{\ell}\in\mathbb{R}^{N\times3}$.

The discrete head biases are initialized to $\log(\hat{p}_m)$, where $\hat{p}_m \!\in\!\{\hat{p}_{\text{pump}},\,\hat{p}_{\text{idle}},\,\hat{p}_{\text{turb}}\}$ is the mode distribution from empirical optimal solutions. This makes the initial softmax match the prior and reduces infeasible mode combinations~\cite{lin2017focal}.

\subsection{Differentiable Integer Relaxation}
\label{sec:ste}

The differentiable integer relaxation block in Fig.~\ref{fig:midpc} maps mode logits to hard modes while preserving gradients.

\paragraph{Training with Gumbel-Softmax}
During training, mode logits $\boldsymbol{\ell}_t\in\mathbb{R}^3$ are discretized via the Gumbel-Softmax straight-through estimator~\cite{jang2016categorical}. Given i.i.d.\ Gumbel noise $g_i$, the soft probabilities are
\begin{equation}
\tilde{s}_{t,i} = \frac{\exp\!\big((\ell_{t,i}+g_i)/\tau\big)}
                       {\sum_{j=1}^{3}\exp\!\big((\ell_{t,j}+g_j)/\tau\big)},
\label{eq:gumbel}
\end{equation}
where $\tau>0$ is the temperature coefficient. The straight-through estimator~\cite{bengio2013estimating} applies argmax in the forward pass to obtain a hard one-hot $\bar{\boldsymbol{s}}_t\in \{0,1 \}^3$, while routing backward gradients through $\tilde{\boldsymbol{s}}_t\in (0,1)^3$. The scalar mode is
\begin{equation}
m_t = \bar{\boldsymbol{s}}_t^\top [-1,\,0,\,1]^\top.
\label{eq:mode_scalar}
\end{equation}
At inference, Gumbel noise is removed and modes are selected deterministically via $\bar{\boldsymbol{s}}_t = \text{one\_hot}(\arg\max_i\,\ell_{t,i})$.

\begin{figure}[t]
    \centering
    \includegraphics[width=\columnwidth]{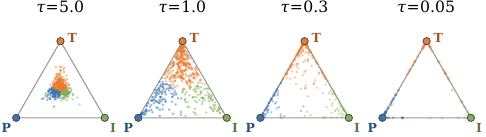}
    \vspace{-0.5cm}
    \caption{Gumbel-Softmax samples on the probability simplex for decreasing temperature coefficient~$\tau$. Vertices correspond to pump~(P), idle~(I), and turbine~(T). Lower~$\tau$ concentrates samples near the vertices.}
    \label{fig:gumbel_simplex}
\end{figure}

\paragraph{Temperature annealing}
The temperature $\tau$ controls exploration versus commitment~\cite{maddison2016concrete}. We keep $\tau\!=\!\tau_0$ during warm-up and then decrease it exponentially toward $\tau_{\text{end}}$:
\begin{equation}
\tau(e)=\begin{cases}
\tau_0, & e < e_w,\\[2pt]
\tau_0\!\left(\dfrac{\tau_{\text{end}}}{\tau_0}\right)^{\!(e-e_w)/(E-e_w)}, & \text{otherwise},
\end{cases}
\label{eq:tau_schedule}
\end{equation}
where $E$ is the total number of epochs and $e_w$ is the warm-up epochs. As illustrated in Fig.~\ref{fig:gumbel_simplex}, high~$\tau$ in warm-up encourages exploration, while low~$\tau$ at the end yields near-hard decisions and narrows the training-inference gap.
\subsection{Differentiable Simulator}
\label{sec:dynamics}

The differentiable simulator block in Fig.~\ref{fig:midpc} unrolls the nonlinear dynamics while preserving end-to-end gradients.

\begin{figure}[t]
    \centering
    \includegraphics[width=\columnwidth]{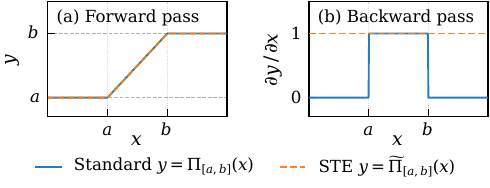}
     \vspace{-0.6cm}
    \caption{Standard clamp vs.\ straight-through (STE) clamp. (a)~Both produce identical forward outputs. (b)~The standard clamp zeros the gradient outside $[a,b]$, while the STE clamp preserves unit gradient everywhere.}
    \label{fig:ste_clamp}
\end{figure}

\paragraph{Straight-through clamp}
A standard clamp $\Pi_{[a,b]}(x):=\min(\max(x,a),b)$ zeros the gradient whenever a state saturates a bound,
\begin{equation}
\frac{\partial\,\Pi_{[a,b]}(x)}{\partial x} = 1(a < x < b),
\label{eq:hard_clamp_grad}
\end{equation}
where $1(\cdot)\in\{0,1\}$ is the indicator function, cutting off the training signal precisely when the policy is out of bounds.
We replace all projections with a straight-through clamp~\cite{bengio2013estimating}: identical to $\Pi_{[a,b]}$ in the forward pass, but with gradient defined as
\begin{equation}
\frac{\partial\,\tilde{\Pi}_{[a,b]}(x)}{\partial x} := 1,
\label{eq:ste_clamp}
\end{equation}
so when gradients flow through saturated states, the policy continues to receive a training signal at the boundary, as in Fig.~\ref{fig:ste_clamp}.
The projection also ensures the profit surrogate is computed over physically feasible trajectories; without it, the policy could exploit arbitrage along infeasible states, producing a misleading training signal (see Section~\ref{sec:loss}).

\paragraph{Power scaling}
The normalized continuous outputs $\boldsymbol{u}_c$ are scaled to the feasible power vector $\mathbf{p}$ using head-dependent bounds and mode indicators. For turbine mode,
\begin{equation}
p_t^T = \big(p_{\min}^T(h_t) + u_{c,t}^T\,(p_{\max}^T(h_t)-p_{\min}^T(h_t))\big)[m_t]_+.
\label{eq:power_scale}
\end{equation}
The pump expression is analogous, using $[-m_t]_+$.

\paragraph{Parallel rollout}
Let $\mathbf{x} := \mathbf{v}$ denote the state variable (reservoir volume) and $\mathbf{s} := \mathbf{h}$ the coupling variable (head).
Given control variable $\mathbf{p}$ (power), the flow rate yields $\mathbf{q}=f^{\text{UPC}}(\mathbf{p},\mathbf{s},\mathbf{m})$ as in~\eqref{eq:upc}, the volume trajectory $\mathbf{v}$ is
\begin{equation}
\mathbf{v} = v^{\text{init}}\,\mathbf{1}^{\top} + \Delta t\,\mathbf{q}\,\mathbf{M}^{\top},
\label{eq:vol_cumsum}
\end{equation}
where $\mathbf{M}\in\mathbb{R}^{T\times T}$ is lower-triangular ($M_{ij}:=\mathbf{1}[j\le i]$), so $(\mathbf{q}\mathbf{M}^\top)_t = \sum_{k=1}^{t}q_k$ accumulates flow causally, recovering the sequential update $v_t = v_{t-1}+\Delta t\,q_t$ from~\eqref{eq:vol_dyn}. Let $f$ denote the dynamics and $g$ the states coupling function:
\begin{equation}
f(\boldsymbol{u}_c,\mathbf{m},\mathbf{s}) := v^{\text{init}}\mathbf{1}^\top + \Delta t\,f^{\text{UPC}}(\mathbf{p},\mathbf{s},\mathbf{m})\mathbf{M}^\top.
\end{equation}
\begin{equation}
g(\mathbf{x}) := (f^{\text{vol}})^{-1}(\mathbf{x}).
\end{equation}
Here $g$ recovers hydraulic head from volume via~\eqref{eq:vol_head}. A natural baseline is the sequential rollout as shown in Algorithm~\ref{alg:step_rollout}, which propagates the clamped state from step $t$ to step $t{+}1$. Although exact, backpropagation through this $N$-step causal chain is analogous to backpropagation through time (BPTT) in recurrent networks, where gradient magnitudes decay exponentially with horizon length~\cite{list2025differentiability,xu2022accelerated}. The parallel rollout shown in Algorithm~\ref{alg:simulator} replaces the sequential loop with a two-pass fixed-point estimate: Pass~1 evaluates all time steps in parallel using a frozen initial state, and Pass~2 refines with the recovered coupling variable. This breaks the $N$-step dependency chain, trading some simulator fidelity for stronger gradient signal. 

\begin{algorithm}[t]
\footnotesize
\caption{Sequential Rollout Simulator}
\label{alg:step_rollout}
\begin{algorithmic}
\Require{Initial condition $x_0=v^{\text{init}},\; s_0=h^{\text{init}}$; control input $\boldsymbol{u}_c$; mode indicators $\mathbf{m}$; STE projection clamp $\tilde{\Pi}$}
\State{\textbf{for} $t = 1,\dots,N$ \textbf{do}}
\State{\In{$\hat{x}_t \leftarrow f(u_{c,t},\,m_t,\,s_{t-1})$}} \algocomment{one-step dynamics}
\State{\In{$x_t \leftarrow \tilde{\Pi}(\hat{x}_t)$}} \algocomment{STE clamp}
\State{\In{$s_t \leftarrow g(x_t)$}} \algocomment{recover coupling variable}
\State{\textbf{end for}}
\AlgReturn{$(\mathbf{x},\,\hat{\mathbf{x}},\,\mathbf{s})$} \algocomment{projected, raw, coupling variables}
\end{algorithmic}
\end{algorithm}

\begin{algorithm}[t]
\footnotesize
\caption{Parallel Rollout Simulator}
\label{alg:simulator}
\begin{algorithmic}
\Require{Initial condition $\hat{s}_0=h^{\text{init}}$; control input $\boldsymbol{u}_c$; mode indicators $\mathbf{m}$; STE projection clamp $\tilde{\Pi}$}
\Statex[\textit{--- Pass 1: frozen initial state ---}]
\State{$\hat{\mathbf{x}} \leftarrow f(\boldsymbol{u}_c,\,\mathbf{m},\,\hat{s}_0\,\mathbf{1}^\top)$} \algocomment{batch dynamics with fixed state}
\State{$\mathbf{x} \leftarrow \tilde{\Pi}(\hat{\mathbf{x}})$} \algocomment{STE clamp to feasible set}
\State{$\mathbf{s} \leftarrow g(\mathbf{x})$} \algocomment{recover coupling variable}
\Statex[\textit{--- Pass 2: refined state ---}]
\State{$\hat{\mathbf{x}} \leftarrow f(\boldsymbol{u}_c,\,\mathbf{m},\,\mathbf{s})$} \algocomment{re-evaluate with recovered state}
\State{$\mathbf{x} \leftarrow \tilde{\Pi}(\hat{\mathbf{x}})$} \algocomment{STE clamp}
\State{$\mathbf{s} \leftarrow g(\mathbf{x})$} \algocomment{output map}
\AlgReturn{$(\mathbf{x},\,\hat{\mathbf{x}},\,\mathbf{s})$} \algocomment{projected, raw, coupling variables}
\end{algorithmic}
\end{algorithm}

\subsection{Loss Function Design}
\label{sec:loss}

The loss combines profit and feasibility penalties:
\begin{equation}
\mathcal{L} = -L^{\text{ex}} + \mathcal{L}^{\text{feas}}.
\label{eq:total_loss}
\end{equation}

\paragraph{Ex-post profit surrogate}
Let $p_t^{\text{opt}}$ denote scheduled net power from~\eqref{eq:power_scale} and $p_t^{\text{sim}}$ the realized net power from the differentiable simulator. The ex-post profit is
\begin{equation}
\begin{aligned}
L^{\text{ex}} ={}&
\underbrace{\sum_{t=1}^{N}\!p_t^{\text{sim}}\lambda_t^{\text{DA}}}_{\text{revenue}}
\;-\;\underbrace{C_{\text{op}}\sum_{t=1}^{N}\!(p_t^{\text{sim}})^2}_{\text{operational cost}} \\
&-\;\underbrace{\omega_{\text{SI}}\sum_{t=1}^{N}\!(p_t^{\text{sim}}-p_t^{\text{opt}})\,\lambda_t^{\text{SI}}}_{\text{system imbalance penalty}}\!-\!\underbrace{\omega_{\text{TV}}\alpha\bar{\lambda}[v_{\text{low},N}-v_{\text{low}}^{\text{target}}]_+}_{\text{target volume penalty}},
\end{aligned}
\label{eq:expost_profit}
\end{equation}
where $\lambda_t^{\text{SI}}$ is the system-imbalance price ($2\lambda_t^{\text{DA}}$ for shortages, $0.5\lambda_t^{\text{DA}}$ for surpluses), $\bar{\lambda}$ is the median day-ahead price, $\alpha$ converts volume to energy, $\omega_{\text{SI}}$ and $\omega_{\text{TV}}$ are penalty weights, and $v_{\text{low},N}$ is the final reservoir volume. The last two terms are the market costs of Section~\ref{sec:problem}: schedule deviations and excess of $v_{\text{low},N}$ over $v_{\text{low}}^{\text{target}}$.

\paragraph{Feasibility penalties}

The soft feasibility term penalizes violations of the raw volume $v_t^{\text{raw}}$ and head $h_t^{\text{raw}}$, i.e., the components of $\hat{\mathbf{x}}$ and $g(\hat{\mathbf{x}})$ in Algorithm~\ref{alg:simulator}:
\begin{equation}
\begin{aligned}
\mathcal{L}^{\text{feas}} ={}&
\kappa_v^-\!\sum_{t=1}^{N}[-v_t^{\text{raw}}]_+
+ \kappa_v^+\!\sum_{t=1}^{N}[v_t^{\text{raw}}\!-\!v_{\max}]_+ \\
&+ \kappa_h^-\!\sum_{t=1}^{N}[h_{\min}\!-\!h_t^{\text{raw}}]_+
+ \kappa_h^+\!\sum_{t=1}^{N}[h_t^{\text{raw}}\!-\!h_{\max}]_+,
\end{aligned}
\label{eq:feas_loss}
\end{equation}
where $\kappa_v^{\pm}$ and $\kappa_h^{\pm}$ are penalty weights. Together, the STE clamp and feasibility penalties form a two-sided training signal: the clamp keeps the profit surrogate~\eqref{eq:expost_profit} honest by enforcing feasibility in the forward pass, while the penalties push raw states back inside bounds in the backward pass. 

\subsection{Training Strategy}
\label{sec:training}

Algorithm~\ref{alg:training} summarizes the policy optimization procedure. Each epoch samples mini-batches from $\mathcal{P}_{\boldsymbol{\xi}}$, rolls out the differentiable pipeline, and updates $\theta$ by backpropagation with gradient clipping. The temperature schedule is given by~\eqref{eq:tau_schedule}, while
hyperparameters are provided in Section~\ref{sec:results}.

\begin{algorithm}[t]
\footnotesize
\caption{MI-DPC Policy Optimizatoin}
\label{alg:training}
\begin{algorithmic}
\Require{Neural control policy $\pi_\theta$; parametric distribution $\mathcal{P}_{\boldsymbol{\xi}}$;  number of epochs $E$; temperature schedule $\tau(e)$}
\State{\textbf{for} $e = 1,\dots,E$ \textbf{do}}
\State{\In{$\tau \leftarrow \tau(e)$}} \algocomment{temp. annealing}
\State{\In{\textbf{for} each mini-batch $\{\boldsymbol{\xi}^{(i)}\}\sim\mathcal{P}_{\boldsymbol{\xi}}$ \textbf{do}}}
\State{\Inn{$(\boldsymbol{u}_c, \boldsymbol{\ell}) \leftarrow \pi_\theta(\boldsymbol{\xi})$} \algocomment{neural policy}}
\State{\Inn{$\boldsymbol{m} \leftarrow \text{GumbelSTE}(\boldsymbol{\ell},\,\tau)$} \algocomment{integer relaxation}}
\State{\Inn{$(\mathbf{x},\,\hat{\mathbf{x}} ,\, \mathbf{s}) \leftarrow \text{Simulate}(\boldsymbol{u}_c, \boldsymbol{m})$} \algocomment{Alg.~\ref{alg:simulator}}}
\State{\Inn{$\mathcal{L} \leftarrow -L^{\text{ex}} + \mathcal{L}^{\text{feas}}$} \algocomment{Eqs.~\eqref{eq:expost_profit}--\eqref{eq:feas_loss}}}
\State{\Inn{$\nabla_\theta\mathcal{L} \leftarrow \text{backprop}(\mathcal{L})$}}
\State{\Inn{Clip $\|\nabla_\theta\mathcal{L}\|_\infty$ and update $\theta$ via AdamW optimizer}}
\State{\In{\textbf{end for}}}
\State{\textbf{end for}}
\AlgReturn{Trained policy $\pi_\theta$}
\end{algorithmic}
\end{algorithm}
\section{CASE STUDY}
\label{sec:results}


The case study uses a representative Belgian UPHES plant with hydraulic head $h\!\in\![50,\,99]$~m and reservoir capacity $v_{\max}\!=\!588{,}000$~m$^3$; UPCs are from laboratory measurements on a reduced-scale Francis pump-turbine~\cite{mercier2017provision,smartwater_multitel_2022}. Day-ahead prices are from the Belgian Elia TSO~\cite{EliaDayAheadReferencePrice}, with operational cost at $0.4$~EUR/MW$^2$ and imbalance multipliers of $2\times$ (shortage) and $0.5\times$ (surplus). The Transformer policy uses mode-logit bias $\log\hat{p}$ with $\hat{p}\!=\![0.40,\,0.15,\,0.45]$ derived from optimal MIQP-PW mode distributions. Training uses AdamW optimizer, batch~32, 25~epochs on 10{,}000 scenarios sampled from a distribution fitted to 2024 Elia prices, excluding the evaluation set. Key hyperparameters: $\tau_0{=}10\to\tau_{\text{end}}{=}0.08$ ($e_w{=}0.35E$, i.e.\ 9 of 25 epochs); $\kappa_v{=}\kappa_h{=}50$; $\omega_{\text{SI}}{=}\omega_{\text{TV}}{=}1$. The 19 held-out evaluation dates are selected by K-medoids clustering on 2024 Elia prices.

The MI-DPC pipeline is implemented using the NeuroMANCER library~\cite{Neuromancer2023}, an open-source differentiable programming framework for parametric constrained optimization built on PyTorch~2.9.1 (CUDA~13.0). MIQP baselines are solved with Gurobi~13.0.0.
All experiments run on an Intel Core Ultra~9 275HX with 32~GB RAM and an NVIDIA GeForce RTX~5070 Laptop GPU.

\subsection{Benchmark Comparison}
\label{sec:benchmark}

Table~\ref{tab:results} shows the performance of MI-DPC against MIQP-GL and MIQP-PW on the 19 held-out benchmark days. All methods are scored by the same ex-post profit~\eqref{eq:expost_profit} under the exact nonlinear simulator: the MIQP baselines optimize~\eqref{eq:objective} under approximate dynamics, so their schedules $p_t^{\text{opt}}$ deviate from the realized $p_t^{\text{sim}}$ and incur imbalance and target-volume penalties, whereas MI-DPC trains on~\eqref{eq:expost_profit} directly. MI-DPC statistics are mean and standard deviation over 47 seeds; the MIQP baselines are deterministic, with the optimal gap set at 1\% and maximum solution time at 1 hour.

\begin{table}[h]
\centering
\caption{Mean ex-post profit (EUR/day) across 19 benchmark days. }
\label{tab:results}
\begin{tabular}{@{}lccc@{}}
\toprule
Method & Profit (EUR/day) & Training time & Inference time \\
\midrule
MIQP-GL & 1{,}997 & --- & 1.91\,s \\
MIQP-PW & 2{,}530 & --- & 918.89\,s \\
\textbf{MI-DPC} & $\mathbf{2{,}489 \pm 71}$ & $\mathbf{161}$ \textbf{s}  & $\mathbf{4.9}$\,\textbf{ms} \\
\bottomrule
\end{tabular}
\end{table}

MI-DPC achieves $2{,}489\!\pm\!71$~EUR/day, a 24.6\% improvement over MIQP-GL and within 1.6\% of MIQP-PW. Its key advantage is online speed: 4.9~ms per day ($390\times$ faster than MIQP-GL and over five orders of magnitude faster than MIQP-PW), making it the only method compatible with real-time re-dispatch as market conditions update. MI-DPC requires 161~s of one-time offline training, whereas the MIQP baselines solve from scratch at every invocation.

\begin{figure}[t]
    \centering
    \includegraphics[width=\columnwidth]{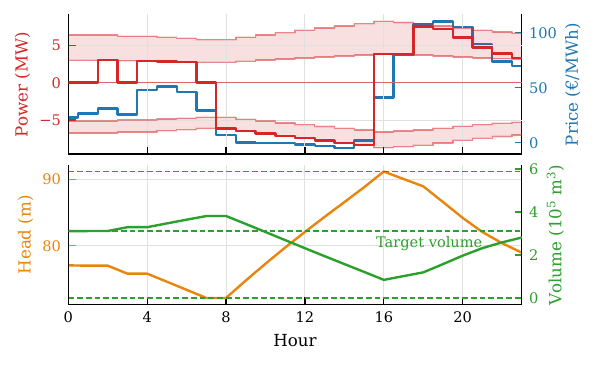}
    \vspace{-2em}
    \caption{MI-DPC converged schedule for a representative day. \textit{Top:} power dispatch (red, MW) and day-ahead price (blue, €/MWh); shaded red bands indicate the feasible turbine (positive) and pump (negative) power regions. \textit{Bottom:} hydraulic head (orange) and reservoir volume (green); dashed lines mark the volume bounds and end-of-day target.}
    \label{fig:epoch_dispatch}
\end{figure}

Fig.~\ref{fig:epoch_dispatch} shows the converged MI-DPC schedule for a representative day. The dispatch lies strictly within the feasible turbine and pump power regions at every hour, confirming constraint satisfaction. The policy concentrates on pumping during low-price hours and turbining during peak-price hours, exploiting day-ahead price arbitrage, while head and volume stay within bounds and end near the target.

\subsection{Ablation Study}
\label{sec:ablation}

\begin{figure*}[t]
    \centering
    \includegraphics[width=\textwidth]{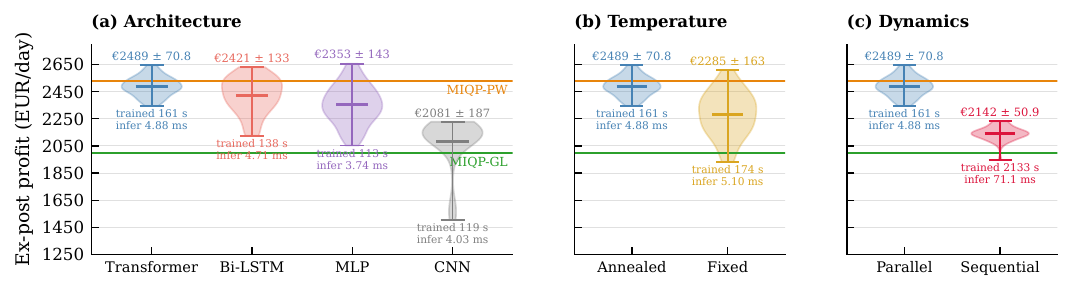}
    \caption{Ablation study: per-seed ex-post profit distributions (47 seeds each). Violin bodies show KDE density; horizontal lines mark the median and interquartile range. The dashed horizontal line indicates the MIQP-PW mean. (a)~Architecture: the Transformer dominates, with MLP and Bi-LSTM as closest competitors. (b)~Temperature schedule: annealing reduces cross-seed variance and improves mean profit. (c)~Training dynamics: the parallel simulator substantially outperforms the sequential rollout simulator.}
    \label{fig:ablation_violins}
\end{figure*}

We ablate three design choices (policy architecture, temperature schedule, and training simulator), each evaluated over 47 random seeds on the same 19 benchmark days. Unless stated otherwise, all runs use the Transformer backbone, annealed temperature, and parallel dynamics from Table~\ref{tab:results}. Fig.~\ref{fig:ablation_violins} shows the per-seed profit distributions for all three ablations.

\paragraph{Policy architecture}
The Transformer achieves the highest mean ex-post profit ($2{,}489\!\pm\!71$~EUR/day), followed by Bi-LSTM ($2{,}421\!\pm\!133$), MLP ($2{,}353\!\pm\!143$), and CNN ($2{,}081\!\pm\!187$); see Fig.~\ref{fig:ablation_violins}(a). Self-attention captures the long-range price dependencies due to arbitrage.

\paragraph{Temperature schedule}
The two-stage annealing schedule~\eqref{eq:tau_schedule} outperforms fixed temperature ($\tau\!=\!0.08$ throughout): $2{,}489\!\pm\!71$ vs.\ $2{,}285\!\pm\!163$~EUR/day (Fig.~\ref{fig:ablation_violins}(b)). Annealing also cuts cross-seed standard deviation (163 to 71~EUR/day): warm-up regularizes the search.

\paragraph{Training simulator}
The parallel rollout (Algorithm~\ref{alg:simulator}) dominates the sequential rollout (Algorithm~\ref{alg:step_rollout}): $2{,}489\!\pm\!71$ vs.\ $2{,}142\!\pm\!51$~EUR/day (Fig.~\ref{fig:ablation_violins}(c)). Backpropagation through the $N$-step causal chain in Algorithm~\ref{alg:step_rollout} is equivalent to BPTT, causing vanishing-gradient effects that degrade policy learning. The sequential rollout is also $12\times$ slower to train ($2{,}133$ vs.\ $161$~s).


\section{CONCLUSION}

This paper extended MI-DPC to nonconvex polynomial dynamics and multi-modal discrete decisions for UPHES scheduling via a parallel differentiable simulator, a Transformer encoder, and a two-stage temperature schedule. The MI-DPC framework achieves ex-post profit within 1.6\% of a piecewise MIQP baseline at millisecond inference times, a five orders of magnitude acceleration compatible with real-time re-dispatch. Ablations show that the Transformer captures long-range arbitrage dependencies, temperature annealing prevents premature mode commitment, and the parallel simulator preserves gradient magnitude.

Future work includes addressing uncertainties arising from plant-model mismatch and price forecasts.
From a theoretical perspective, we aim to derive rigorous feasibility guarantees for mixed-integer optimal control problems with nonconvex polynomial dynamics via tractable safety filters. 


\bibliographystyle{IEEEtran}
\bibliography{reference}

\end{document}